\documentclass[%
 reprint,
superscriptaddress,
showpacs,preprintnumbers,
 amsmath,amssymb,
 aps,
pra,
]{revtex4-1}
\usepackage[dvipsnames]{xcolor}
\usepackage[colorlinks=true,bookmarks=false,linkcolor=NavyBlue,urlcolor=NavyBlue,citecolor=NavyBlue,breaklinks]{hyperref}
\usepackage{amsmath}
\usepackage{amsfonts}
\usepackage{graphicx}
\usepackage{color}
\usepackage{braket}
\usepackage{graphicx}
\usepackage{dcolumn}
\usepackage{bm}

\begin{document}

\title{Newton's second law in spin-orbit torque}

\author{Cong Son Ho}
 \email{elehcs@nus.edu.sg}
 \affiliation{
Department of Electrical and Computer Engineering, National University of
Singapore, 4 Engineering Drive 3, Singapore 117576.}

\author{Seng Ghee Tan}
\affiliation{
Department of Physics, National Taiwan University, Taipei 10617 Taiwan.}
\affiliation{
Department of Physics, Chinese Cultural University, 55 Hwa-Kang Road, Yang-Ming-Shan, Taipei 11114, Taiwan.}
\author{Shun-Qing Shen}
\affiliation{
Department of Physics, The University of Hong Kong, Pokfulam Road, Hong Kong, China.}
\author{Mansoor B. A. Jalil}%
 \email{elembaj@nus.edu.sg}
\affiliation{
Department of Electrical and Computer Engineering, National University of
Singapore, 4 Engineering Drive 3, Singapore 117576.}

\date{\today}

\begin{abstract}
Spin-orbit torque (SOT) refers to the excitation of magnetization dynamics via spin-orbit coupling under the application of a charged current. In this work, we introduce a simple and intuitive description of the SOT in terms of spin force. In Rashba spin-orbit coupling system,  the damping-like SOT can be expressed as ${\mathbf T}^\mathrm{so}={\mathbf R}_c\times {\mathbf F}^{{\mathrm {so}}}$, in analogy to the classical torque-force relation, where $R_c$ is the effective radius characterizing the Rashba splitting in the momentum space. As a consequence, the  magnetic energy is transferred to the conduction electrons, which dissipates through Joule heating at a rate of $({\mathbf j}_e\cdot {\mathbf F}^{\mathrm {so}})$, with $j_e$ being the applied current. Finally, we propose an experimental verification of our findings via measurement of the anisotropic magnetoresistance effect.
\end{abstract}

\maketitle


\section{Introduction}

In classical physics, force is a quantity that can change the motion of an
object. If the object is in rotation motion, torque is introduced to conveniently describe
the driving mechanism instead of the force. In general, the force ($\mathbf{F}$) and the torque ($\mathbf{T}$)
are related by $\mathbf{T}=\mathbf{R}\times \mathbf{F}$, where $\mathbf{R}$ is the radius vector. In other words, the torque is obtained once the force and radius vector are known.

In quantum mechanics, the force concept is less well-defined,
and there are fewer application of force equation in describing physical
phenomena. Instead, several approaches such as the Schrodinger equation have taken the primary role in describing the quantum state and the evolution of  physical systems. However, in
some contexts, the force concept is still useful in the sense that it can provide an intuitive description
of quantum effects. For example, in systems with spin-orbit coupling, the spin-orbit field mimics an effective magnetic field
that induces a Lorentz-like force \cite{Shen:prl05, Ho:epl14,Tan:jpsj13}, that provides an explicit connection
between the spin dynamics and the motion of electron. This linkage gives an underlying intuitive picture of  various spin transport
effects. For example,
Zitterbewegung (jitter motion) is the trembling motion of electron resulting from
the interference between positive and negative energy branches (corresponding to
up-spin and down-spin, respectively) \cite{Schlie:prb06}. In the spin force picture \cite{Shen:prl05}, it has been shown that opposite spins experience opposite forces, thus an electron with precessing spin will experience an oscillating force, leading to its trembling motion. Similarly, a rf current is shown to generate oscillating spin-force on electron, which leads to current induced electron spin resonance \cite{ESR:prb12}. Another phenomenon is the spin Hall effect (SHE) which refers to the generation of a transverse pure spin current by application of a longitudinal electric field \cite{Sino:prl04,Mura:sci03}. The SHE had been explained by concepts such as anomalous velocity \cite{Mura:sci03} and/or momentum-dependent spin polarization \cite{Sino:prl04}. At the same time, it can also be intuitively described by the spin
force picture, in which a longitudinal force due to an electric field induces a transverse spin Hall current \cite{Tan:jpsj13,Ho:epl14}, in an analogous way to the classical Hall effect. In the presence of crystal field and scattering,  an extension of the Drude model can be derived via spin-force in describing the SHE \cite{PhysRevLett.99.206601,PhysRevB.80.153105}. Moreover,  the disappearance of the intrinsic SHE in the presence of impurity scattering in Rashba system may also be explained using the spin force picture\cite{Ada05}, an outcome which can be confirmed more rigorously via vertex corrections \cite{Inoue:prb04}.

At the same time, the concept of torque has been widely used
in the fields of magnetism and spintronics to describe the dynamics of magnetic moment and electron spin \cite{Ralph:jmmm08,Manchon:prb08,Kure:nat14}. In systems with dynamic magnetic texture, electrons experience Lorentz-like force due to emergent effective electromagnetic field \cite{Zhang:prl05,Barn:prl07,Kim:prl12, Ho:NJP15}. In such systems, the effective electric field is associated with the so-called spin-motive force, that can drive spin currents, which in turn, exerts a torque on the magnetic texture through the spin-transfer torque mechanism \cite{Zhang:prl05,Kim:prl12}. On the other hand, in a static and uniform magnetic system, the spin-motive force vanishes and so does the spin-transfer torque. Instead, the magnetization will experience a spin-orbit torque (SOT) which originates  from the current-induced spin polarization \cite{Manchon:prb08,Kure:nat14}. Thus, a natural question that arises is whether the spin-force and the spin-torque are counterparts of each other, and if they are related in a similar way as in classical mechanics.

On the other hand, another issue to be clarified is concerning the energy dissipation induced by the damping torque. It has been well-known that the intrinsic Gilbert damping causes the magnetic energy to dissipate through itinerant quasiparticles \cite{Blank}, e.g., phonon \cite{Bra:prl08}, magnon \cite{Lindner:prb03}. However, it has recently been shown that spin-orbit torque can also induce damping \cite{Kure:nat14}, which has a different mechanism from the Gilbert damping and thus may have a different channel of energy dissipation. Thus, another question that arises is to ascertain where the energy loss induced by the damping SOT gets transferred to.

To address the above two issues, we will consider the spin-orbit torque induced in Rashba SOC
system. We show that the spin-force can indeed be connected to the SOT. However,
since the force represents a dynamical process, i.e., when the system is not under steady state, the SOT is generated in a
dynamical state. This connection is consistent since the
anti-damping torque originates from the Berry phase, which is similar to that of the SHE. Moreover, from the force-torque equation, we will show that the magnetic energy loss due to damping SOT is transferred to conduction electrons, which finally dissipates in the form of Joule heating. Finally, we propose an experimental verification of our findings via measurement of the anisotropic magnetoresistance effect.
\section{Theory}
	
We begin with the model Hamiltonian of two-dimensional electron gas (2DEG) coupled to ferromagnet in the presence of the Rashba spin-orbit coupling \cite{Rashba}, which is given by
\begin{equation}\label{hal}
\mathcal{H}=\frac{{\mathbf p}^2}{2m_{\mathrm e}}+{\alpha}({\mathbf k}\times {{\mathbf z}})\cdot \mathbf{\sigma} -J_\mathrm {ex}{\mathbf M}\cdot\mathbf{\sigma} +V({\mathbf r}).
\end{equation}
where ${\mathbf p}=\hbar (k_x,k_y)$ and $\mathbf{\sigma}=(\sigma_x,\sigma_y,\sigma_z)$ are the momentum and the Pauli matrix, respectively. ${{\mathbf z}}$ is the unit vector perpendicular to the plane, $\alpha$ is the Rashba parameter, $J_\mathrm {ex}>0$ is the exchange coupling between the electron spin and the magnetization, and ${\mathbf M}$ is the magnetization unit vector.
The last term $V({\mathbf r})=e{\mathbf E}\cdot{\mathbf r}$ is the applied electric field.

The spin dynamics is described by the Ehrenfest theorem
derived from (\ref{hal}) as \cite{Manchon:prb08,manchon2018current}
\begin{equation}\label{eq2}
\frac{d{\mathbf S}}{dt} =  -\nabla\cdot \mathcal{J}_s+\frac{1}{\hbar}\braket{{\mathbf B}^{\mathrm {so}}\times\mathbf{\sigma}}-\frac{J_\mathrm {ex}}{\hbar} {\mathbf M}\times{\mathbf S},
\end{equation}
where ${\mathbf S}=\braket{{\mathbf\sigma}}$ is the spin polarization, $\mathcal{J}_s$ is the spin current, and ${\mathbf B}^{\mathrm {so}}=\alpha({\mathbf k}\times {{\mathbf z}})$. The braket $\langle \dots\rangle$ represents the expectation value taken over energy eigenstates of the system, $\braket{\mathcal{O}}=\sum_{n\mathbf{k}}\braket{\psi_{n\mathbf{k}}|\mathcal{O}|\psi_{n\mathbf{k}}}f_{n\mathbf{k}}$, with $\psi_{n\mathbf{k}}$ and $f_{n\mathbf{k}}$ being the eigen-state and the distribution function corresponding to eigen-energy $E_{n\mathbf{k}}$, respectively. At the same time, the magnetization dynamics is described by following equation

\begin{equation}\label{LLG1}
\frac{d{\mathbf M}}{dt}=\left[\frac{d{\mathbf M}}{dt}\right]_\mathrm{LLG}+\frac{J_\mathrm {ex}}{\hbar} {\mathbf M}\times{\mathbf S},
\end{equation}
where $[d{\mathbf M}/dt]_\mathrm{LLG}=-\gamma {\mathbf M}\times {\mathbf H}_\mathrm{0}+\gamma_G {\mathbf M}\times \frac{d{\mathbf M}}{dt}$ is the conventional LLG equation, in which ${\mathbf H}_\mathrm{0}$ includes anisotropy fields and external magnetic field, $\gamma_G$ is the intrinsic Gilbert damping constant, and $\gamma$ is the gyromagnetic ratio. 

In the above, the second term in Eq. (\ref{LLG1})
represents the coupling between the spin and magnetization dynamics and it is usually referred as the
spin-orbit torque acting on the magnetization \cite{manchon2018current,haney2008current}, which is related to the spin dynamics via Eq.\ref{eq2}. We note here that, in general, the SOT can stem from two origins \cite{brataas2012current, PhysRevB.94.104419, manchon2018current}: spin Hall effect (SHE) , and inverse spin galvanic effect (iSGE). In the former case, the generated spin current via bulk SHE in the non-magnetic (NM) layer is injected into the coupled ferromagnetic (FM) layer where the angular momentum of the conduction electron is transferred to the local magnetic moments \cite{brataas2012spin,PhysRevB.66.014407,PhysRevB.94.104420}. This process is modeled by the spin drift-diffusion equation \cite{PhysRevB.71.045123,PhysRevB.87.174411, PhysRevB.94.104420}. The SHE-induced SOT is in analogy to the spin-transfer torque (STT), where the NM with strong SOC plays the role of the pinning (polarizing) FM layer \cite{brataas2012spin,PhysRevB.66.014407,PhysRevB.94.104420}.  On the other hand, the iSGE refers to the current-induced spin polarization, which occurs in 2DEG with interfacial SOC \cite{EDELSTEIN1990233,Tan:arxiv07,PhysRevB.77.214429, Manchon:prb08,Kure:nat14}. As it has been shown, the iSGE-SOT can be understood in terms of current-driven spin texture in the momentum space \cite{Manchon:prb08,Kure:nat14,manchon2018current}. In contrast to the SHE-induced SOT, where the spin polarization is only accumulated at the edges, the spin polarization in iSGE-induced SOT is uniformly generated in the entire conduction channel.  Furthermore, as it has been shown  earlier \cite{PhysRevB.94.104420}, when the NM thickness reduces, the SHE-induced SOT will diminish and vanish in a thin NM film, whereas the iSGE-SOT is almost constant. This can be understood as the result of vanishing bulk SOC , which is the origin of the SHE-SOT, while the interfacial SOC still remain when the system approaches quasi-2DEG. Nevertheless, there are some attempts to include both types of SOT in the same framework by using, for example, Kubo-Bastin formula \cite{PhysRevB.90.174423,PhysRevB.92.064415, PhysRevB.91.014417}, generalized magnetoelectronic circuit theory  \cite{PhysRevB.94.104419}. In previous works \cite{Shen:prl05,Ho:epl14}, the spin-force picture has been used to describe various effects due to the spin texture in the SOC system. Hence, in our work, we restrict ourselves to the second contribution, i.e., SOT induced by iSGE or spin texture in 2DEG Rashba system.

Assuming a uniform magnetization which is the case of single ferromagnet \cite{Manchon:prb08,manchon2018current}, the spin current is also uniformly distributed such that $\nabla\cdot \mathcal{J}_s=0$, from which the spin-torque is obtained from (\ref{eq2}) as

\begin{equation}\label{torque}
{\mathbf T}=\frac{J_\mathrm {ex}}{\hbar} {\mathbf M}\times{\mathbf S} =  \frac{1}{\hbar}\braket{{\mathbf B}^{\mathrm {so}}\times\mathbf{\sigma}}-\frac{d{\mathbf S}}{dt}.
\end{equation}

In general, the central objective in derivation of the SOT is to determine the expectation value of
the spin polarization, which is the solution of equation (\ref{eq2}). When the system is at steady state, the solution can be obtained by setting the left-hand side of Eq. (\ref{eq2}) to zero,  i.e.
${d{\mathbf S}}/{dt}=0$. The steady-state SOT can be derived as ${\mathbf T}={\mathbf T}^{(1)}$, where
\begin{equation}\label{steady}
{\mathbf T}^{(1)}=\frac{1}{\hbar}\langle {\mathbf B}^{\mathrm {so}}\times \mathbf{\sigma} \rangle.
\end{equation}
 The above torque can be calculated by several methods
such as the Boltzmann transport equation \cite{Manchon:prb08}, Kubo formula \cite{LiHang:prb15}, and gauge theory method \cite{Tan:arxiv07,Tan:ann11}. Explicitly, it is given by
${\mathbf T} ^{(1)}=-\gamma({\mathbf M}\times {\mathbf H}^\mathrm{fl})$, which is usually referred to as field-like torque \cite{Tan:arxiv07,Tan:ann11}, where $ {\mathbf H}^\mathrm{fl}=\frac{m_{\mathrm e}\alpha}{e\hbar}({{\mathbf z}}\times {\mathbf j}_e)$, and ${\mathbf j}_e $ is the charge current density. A brief derivation of the field-like torque via gauge theory approach is given in \ref{ap2}.

We recall the intuitive picture of the field-like torque under the steady state as follows\cite{Manchon:prb08,manchon2018current}: an applied electric field in the $x$-direction will induce a
shift of the Fermi circle in the $k_x$ direction in momentum space. Since the Rashba field is locked to the momentum
by ${\mathbf B}^{\mathrm {so}}\propto ({\mathbf k}\times {{\mathbf z}})$, a net shift of the momentum $\Delta k_x$ in the $k_x$-direction will result in a net Rashba
field in the $y$-direction ${\mathbf B}^{\mathrm {so}}\propto \Delta{k}_x{{\mathbf y}} $. This in turn induces a (field-like) torque on
the coupled magnetization. Under steady state, $d{\mathbf k}/dt=0$,
the net momentum shift is given $\Delta  k_x=eE_x \tau/\hbar$, where $\tau$ is the momentum relaxation
time. By the relation between the charge current and momentum shift we obtain $j_e\propto \Delta k$, from which we recover the field-like torque expression given above. Therefore, from this exercise we may conclude that the
field-like torque is induced by non-equilibrium spin polarization under steady
state.

On the other hand, a question that arises as to  what spin torque would be  if the system has not yet reached the 
steady state, i.e., $d{\mathbf k}/dt\ne 0$ and $d{\mathbf S} /dt\ne 0$? In this case, the left-hand side of Eq. (\ref{eq2}) retains and the total spin-torque becomes


\begin{equation}
{\mathbf T}={\mathbf T}^{(1)}+{\mathbf T}^{(2)},
\end{equation}
where the first term 
related to the steady state, which gives rise to the field-like torque ${\mathbf T}^{(1)}$ given by Eq. (\ref{steady}), and the
additional term related to the spin dynamics in dynamical state as
\begin{equation}\label{dyn}
{\mathbf T}^{(2)}=-\frac{d{\mathbf S}}{dt}.
\end{equation}
One may wonder if we can have both torque terms at the same time? Such a situation arises, e.g., when we consider the system at the moment that it almost reaches its steady state
such that the momentum is shifted by $\Delta k\propto eE\tau$, but its time-derivative is still
non-zero, $d{\mathbf k}/dt\ne 0$. We now turn our attention to evaluating the new torque term, i.e., ${\mathbf T}^{(2)}$.

\section{Spin-dependent force}

In spin-orbit coupling system, there is a spin-dependent component of the velocity of electron, which is referred to as the anomalous velocity. As a consequence, the force acting on electron is also associated with the spin operator \cite{Shen:prl05,Ho:epl14}. At the same time, the spin induces a torque on the magnetization as discussed in the previous section. Therefore, it is natural that the spin torque and spin force can be expressed in terms of each other, which we shall explicitly show below. 

Let us first rewrite the Hamiltonian (\ref{hal}) in terms of SU(2) gauge field as following \cite{Ho:epl14,Tan:jpsj13}
\begin{equation}\label{hal2}
\mathcal{H}=\frac{{\mathbf \Pi}^2}{2m_{\mathrm e}}-J_\mathrm {ex}{\mathbf M}\cdot\mathbf{\sigma} +e{\mathbf E}\cdot{\mathbf r}+\mathcal{O}(\alpha^2),
\end{equation}
where $\Pi= \mathbf{p}+e\mathcal{A}$ is the canonical momentum, with ${\mathcal A}=m_{\mathrm e}\alpha/e\hbar ({{\mathbf z}}\times\sigma)$ being the Rashba SU(2) gauge field. The net force acting on electron is calculated as $\mathbf{F}=\frac{d\braket{\Pi}}{dt}=e\mathbf{E}+e {d\braket{\mathcal{A}}}/dt$, where the first term $\dot{\mathbf{p}}=e\mathbf{E}$ being the conventional electric force, and the second term arisen from the spin-dependent gauge field is explicitly read as

\begin{equation}\label{force1}
{\mathbf F}^{\mathrm {so}}=e \frac{d\braket{\mathcal{A}}}{dt}=\frac{m_{\mathrm e} \alpha}{\hbar} \left({{\mathbf z}}\times \frac{d{\mathbf S} }{dt}\right).
\end{equation}
If the spin dynamics equation \eqref{eq2} is substituted into the above, we recover the spin-force equation obtained previously in the absence of the magnetization \cite{Shen:prl05,Ho:epl14}, which can be used to intuitively described the Zitterbewegung \cite{Shen:prl05} and intrinsic spin Hall effect \cite{Ho:epl14}. A similar equation has been introduced in the context of spin resonance under the application of a rf electric field \cite{ESR:prb12}.
 
In our case, the spin dynamics in the right-hand side represents one of the torque components given in Eq. \eqref{torque},  which can be expressed in term of the spin-force by rearranging Eq. \eqref{force1} as
\begin{equation}\label{spin1}
{\mathbf T}^{(2)}=-\frac{d{\mathbf S} }{dt}=\frac{\hbar}{m_{\mathrm e} \alpha} ({{\mathbf z}}\times {\mathbf F}^{\mathrm {so}}),
\end{equation}
The above equation resembles the mechanical torque expression in classical mechanics,
i.e., ${\bf T}^{(2)}={\bf R}_c\times {\bf F}^{\mathrm {so}}$, with ${\bf R}_c=(\hbar/m_e\alpha ){{\bf z}}$ being the effective radius vector. It is obvious that the effective radius is related to the separation of the two Fermi circles of the Rashba bands in momentum space, i.e., $R_c=2(p_{f-}-p_{f+})^{-1}$, where $p_{f\pm}=p_f\mp m_e\alpha/\hbar$ are Fermi momenta of upper and lower bands \cite{Sino:prl04}, respectively. Now, we can see from the torque-force equation a connection between the spin torque ($dS/dt$) in the spin space, the spin separation in the momentum space ($R_c$) and the acceleration of electron due to the spin force in the real space. Once the force is evaluated, the torque can be immediately obtained.

 Now we will explicitly derive the torque induced in the dynamical state by calculating the spin force. Previously, we showed that the force can be derived by a
semi-classical method \cite{Ho:epl14}, i.e., via the Hamilton's equations
\begin{equation}\label{force5}
{\mathbf F}^{\mathrm {so}}=\frac{d}{dt}  \frac{d\epsilon}{\hbar d{\mathbf k}},
\end{equation}
where $\epsilon$ is the eigen-energy of the system. A brief recall of this method is given in \ref{ap1}. In this case, the energy is given by $\epsilon _\pm =\frac{\hbar^2{\mathbf k}^2}{2m_{\mathrm e} }\pm \Omega$,
where $\Omega=|J_\mathrm{ex}{\mathbf M}+\alpha({\mathbf k}\times {{\mathbf z}})|$, and $(\pm )$ are for upper/lower bands, respectively.
Substituting the above into Eq. (\ref{force5}),
the spin force is derived as
\begin{equation}
{\mathbf F}^{\mathrm {so}}=\frac{m_{\mathrm e} \alpha ^2}{\hbar^2\Omega} \left\{\hbar\dot{\mathbf k}-\frac{J_\mathrm {ex}^2}{\Omega^2} ({{\mathbf z}} ̂\times {\mathbf M})\left[({{\mathbf z}}\times {\mathbf M})\cdot
\hbar\dot{\mathbf k}\right]\right\},
\end{equation}
where we have ignored the higher order term than $\alpha^2$. Assuming the strong exchange
limit $J_\mathrm {ex}\gg\alpha k$, we can approximate $\Omega\approx J_\mathrm {ex}$, and noting that $\hbar\dot{\mathbf k}=e{\mathbf E}$, the force equation
can be simplified to
\begin{eqnarray}
{\mathbf F}^{\mathrm {so}}=\eta e {{\mathbf z}}\times \left\{{\mathbf M}\times \left[{\mathbf M}\times ({{\mathbf z}}\times {\mathbf E})\right]\right\},\label{force2}
\end{eqnarray}
where $\eta=\frac{m_{\mathrm e} \alpha^2}{\hbar^2J_\mathrm {ex}}$.

Now we will discuss how interplay of the magnetization and the electric field leads to the generation of the above spin force. Equation (\ref{force2}) can be reduced to ${\mathbf F}^{\mathrm {so}}=\eta e ({\mathbf E}\cdot {\mathbf M}){\mathbf M}$, which means that the spin-force acting on electron is parallel to the magnetization direction. This can be seen as a consequence of the spin-momentum locking in the strong exchange regime. Indeed, when the electron spin aligns with the magnetization,  the canonical momentum is directed in the perpendicular direction, and therefore the force, which is the time-derivative of the momentum, becomes parallel to ${\mathbf M}$ as shown above.  Furthermore, the force strength diminishes as direction of the the applied electrical field and the magnetization becomes perpendicular to one another. In addition, the force is symmetric with respect to the direction of the magnetization.  In the latter part, we will show that the interplay of the magnetization and electric field also leads to the anisotropic magnetoresistance effect.

Now we can immediately obtain the expression of the spin torque in the dynamical state by substituting the expression of the force into Eq. (\ref{spin1}). Explicitly, we have
\begin{equation}
{\mathbf T}^{(2)}=\frac{e\alpha}{\hbar J_\mathrm {ex}} {\mathbf M}\times ({\mathbf M}\times ({\mathbf E}\times {{\mathbf z}})),
\end{equation}
which has the form of damping-like torque. A similar formula has been obtained
previously by Kurebayashi {\it et al}. using a completely different argument \cite{Kure:nat14}, where the damping-like SOT is shown to stem from the Berry phase. Interestingly, we can now see a close connection between the damping SOT and the spin Hall effect \cite{Sino:prl04}. The latter can also be understood in terms of the Berry phase \cite{Shen:prb04}, as well as the spin-force picture \cite{Ho:epl14}. In addition, within the framework of the spin-force, the fact that the damping SOT is zero under steady state as discussed above, parallel the observation of vanishing SHE under steady state \cite{Ada05}.

We have shown that the damping torque is induced in a dynamical state, where the electron spin still precesses about the Rashba field. In the above, we have assumed weak Rashba limit, which allows electron spin to precess in a long interval of time. On the other hand, in a strong Rashba limit, electron spin will rapidly align along the Rashba field, which leads to a steady state, i.e. $(d{\mathbf S}/dt =0)$, and thus the damping torque will also quickly disappear. The vanishing damping torque in the strong Rashba limit is also confirmed via Kubo formula in a previous work \cite{LiHang:prb15}.




\section{Energy transfer}

In the above, we have derived the  damping spin-torque associated with the spin-force. In general, a damping torque will cause the magnetization to dissipate its energy. The remaining question is where the energy dissipates into? In the following, we will show that the magnetization gives up its energy in the form of Joule heating.

{\it Magnetic energy dissipation} - First of all, let us discuss the magnetic energy dissipation induced by the damping process. The energy dissipation rate is given as $\frac{dE_M}{M_s dt}=-{\mathbf H}^\mathrm{eff}\cdot \frac{d{\mathbf {\mathbf M}}}{dt}$, which can be calculated from Eq. (\ref{LLG1}), where ${\mathbf H}^\mathrm{eff}$ is the total effective magnetic field. To determine the role of SOC, we assume that the magnetization is only subjected to spin-orbit torque, i.e., ${\mathbf H}^\mathrm{eff}={\mathbf H}^\mathrm{fl}$.  From the equation (\ref{LLG1}), the magnetic energy dissipation rate is calculated as
\begin{eqnarray}\label{EM1}
\frac{dE_M}{M_s dt}&=&-\gamma_G {\mathbf H}^\mathrm{fl}\cdot({\mathbf M}\times \dot{{\mathbf M}})+{\mathbf H}^\mathrm{fl}\cdot \left(R_c{{\mathbf z}}\times {\mathbf F}^{\mathrm {so}}\right)\nonumber\\
&=&-\gamma \gamma_G |{\mathbf M}\times {\mathbf H}^\mathrm{fl}|^2-\frac{1}{e}\left({\mathbf j}_e\cdot{\mathbf F}^{\mathrm {so}}\right).
\end{eqnarray}
In the above, we have assumed that only first order terms in both Gilbert damping and SOC damping are retained. Obviously, the first term is negative and it represents the conventional dissipation due to the Gilbert damping. At the same time, it is straightforward to verify that the second term is also dissipative. Indeed, by substituting the force equation ({\ref{force2}) into the above, the second term is read as

\begin{eqnarray}\label{EM2}
Q_M=-\eta\rho_0({\mathbf M}\cdot{\mathbf j}_e)^2,
\end{eqnarray}
which is obviously negative, in which we have replaced ${\mathbf E}=\rho_0 {\mathbf j}_e$, with $\rho_0$ being the resistivity. As mentioned in the introduction, while the intrinsic Gilbert damping possibly causes energy relaxation through quasiparticles \cite{Blank} such as phonon \cite{Bra:prl08} and magnon \cite{Lindner:prb03}, the magnetic damping due to spin-orbit torque is generated by a different mechanism and thus induces  energy dissipation through different channel. In the following, we will show that the magnetic energy is transferred to the conduction electron and dissipated through Joule heating. To verify this we now calculate the Joule heating generated in the damping process.

{\it Joule heating} - Let us first recall the Joule heating in classical electrodynamics which is given by ${\mathbf j}_e\cdot {\mathbf E}$. Alternatively, this formula can be formally derived by considering the coupling between charge current ${\mathbf j}_e$ and electric field ${\mathbf E}$, which gives rise to an interaction given by $W={\mathbf j}_e\cdot {\mathbf A}_e$, where $\partial_t{\mathbf A}_e ={\mathbf E}$ with ${\mathbf A}_e$ being the vector gauge potential associated with the electric field. The Joule heating is recovered as $dW/dt=({\mathbf j}_e\cdot {\mathbf F}_e)/e={\mathbf j}_e\cdot {\mathbf E}$, where ${\mathbf F}_e=e\partial_t{\mathbf A}_e=e{\mathbf E}$ is the electric force acting on the electron. 

Similarly, in our case the Rashba spin-orbit coupling can be represented by a gauge field given by ${\mathcal A}=\frac{m_{\mathrm e}\alpha}{e\hbar}({{\mathbf z}}\times\mathbf{\sigma})$.  The coupling between the SOC gauge field and the charge current induces an interaction $W={\mathbf j}_e\cdot{\mathcal A}$ \cite{Baza:prb98,Tan:arxiv07,Tan:ann11}. The Joule heating density $Q_e=dW/dt$ is similarly calculated as
\begin{eqnarray}\label{Joule1}
Q_e={\mathbf j}_e\cdot \frac{d{\braket{{\mathbf A}_e+\mathcal A}}}{dt}
&&={\mathbf j}_e\cdot {\mathbf E}+\frac{m_{\mathrm e}\alpha}{e\hbar}{\mathbf j}_e\cdot\left({{\mathbf z}}\times\frac{d{\mathbf S}}{dt}\right)\nonumber\\
&&={\mathbf j}_e\cdot {\mathbf E}+\frac{1}{e}\left({\mathbf j}_e\cdot{\mathbf F}^{\mathrm {so}}\right),
\end{eqnarray}
where we have used the force equation (\ref{force1}) in the last step. It can be seen that the first term is the conventional Joule heating as discussed early, and the last term represents an additional Joule heating due to the spin-force and the damping SOT, which is exactly the magnetic energy dissipation given in Eq. ({\ref{EM1}).

We note that the Joule heating may have an additional effect on the magnetization dynamics due to thermal fluctuation \cite{Lin:JAP06,Conte:PRL14,PhysRevB.91.134411,Pham:PRA18}. It is well-known that upon passing a current, the temperature of the system may rise due to the Joule heating \cite{Lin:JAP06,Conte:PRL14,Pham:PRA18}, which leads subsequently to secondary effects on the magnetization dynamics. This is as a consequence of the temperature dependence of various parameters of the magnetic material, as well as the effect of thermal fluctuations on the magnetization dynamics as modelled by the stochastic Landau-Lifshitz equation \cite{PhysRevB.91.134411,Li:PRB04,PhysRevB.72.180405}. However, for simplicity in the analytical model, we ignore the thermal effects on the magnetization dynamics, i.e., we do not consider the variation arising from Joule heating.

{\it Anisotropic magnetoresistance (AMR)}- It is interesting that we can relate the above Joule heating to the AMR effect. Indeed, from Eqs. (\ref{EM2}) - (\ref{Joule1}) the total Joule heating is read as $Q_e=\rho_0 j_e^2+\rho_0\eta j_e^2 \cos^2{\theta_{E-M}}$, where the first term is the conventional Joule heating,  and $\theta_{E-M}$ is the relative angle between the magnetization and the electric field. The Joule heating can be represented as $Q=\rho_\mathrm{AMR} j_e^2$, where $\rho_\mathrm{AMR}$ is effective resistivity given by
\begin{equation}
\rho_\mathrm{AMR}/\rho_0=1+\eta \cos^2\theta_{E-M}.
\end{equation}
The above AMR has a quadratic dependence on the Rashba coupling, which is in agreement with previous findings via kinetic theory \cite{AMR4,AMR3,AMR1}. By analyzing this AMR effect, it is possible to experimentally verify the energy transfer and confirm the  spin-force description of the spin-orbit torque.

In summary, in this work we have showed that in SOC system spin force induces a damping torque on the coupled magnetization. The relation between spin force and spin torque is similar to the torque-force equation in classical mechanics, i.e., ${\mathbf T}^{\mathrm {so}} ={\mathbf R}\times {\mathbf F}^{\mathrm {so}}$. Moreover, the damping torque causes the magnetization to transfer its energy to the conduction electrons, which finally dissipates through Joule heating. These findings can be experimentally verified via anisotropic magnetoresistance effect.

\begin{acknowledgments}

The authors would like to acknowledge the MOE Tier I (NUS Grant No. R-263-000-B98-112), MOE Tier II MOE2013-T2-2-125 (NUS Grant No. R-263-000-B10-112) and MOE Tier II MOE2015-T2-1-099 (NUS Grant No. R-380-000-012-112) grants, as well as the National Research Foundation of Singapore under the CRP Programs “Next Generation Spin Torque Memories: From Fundamental Physics to Applications” NRF-CRP12-2013-01 (NUS Grant No. R-263-000-B30-281) for financial support.
\end{acknowledgments}


\appendix

\section{Derivation of field-like torque}\label{ap2}

The Rasbha Hamiltonian can be rewritten as ${\mathcal H}=\frac{1}{2m_{\mathrm e}}({\mathbf p}+e{\mathbf \mathcal{A}})^2 -J_\mathrm {ex}{\mathbf M}\cdot{\mathbf\sigma}+e\mathbf{E}\cdot {\mathbf r}$, where  ${\mathcal A}=m_{\mathrm e}\alpha/e\hbar ({{\mathbf z}}\times\sigma)$ is the Rashba gauge field. The coupling between a charge current and the Rashba gauge field induces interaction energy given by $W={\mathbf j}_e\cdot {\mathcal A}$ \cite{Tan:arxiv07,Tan:ann11}. In a strong exchange regime, the spin
mostly aligns along magnetization direction, i.e., $\braket{\sigma}\approx {{\mathbf M}}$. The interaction energy becomes $W=-m_{\mathrm e}\alpha/e ({{\mathbf z}}\times {\mathbf j}_e)\cdot {{\mathbf M}}$, which represents the interaction between the current and the magnetization \cite{Baza:prb98,Tan:arxiv07,Tan:ann11}.
The current-induced effective magnetic field can be derived by ${\mathbf H}^\mathrm{fl}=-\frac{dW}{M_s d{{\mathbf M}}}$, which is read as

\begin{equation}
{\mathbf H}^\mathrm{fl}=\frac{m_{\mathrm e}\alpha}{e\hbar}({{\mathbf z}}\times {\mathbf j}_e).
\end{equation}
This field induced a (field-like) torque on the magnetization as ${\mathbf T}^{(1)}=-\gamma \left({{\mathbf M}}\times {\mathbf H}^\mathrm{fl}\right)$.

\section{Derivation of classical force}\label{ap1}
Strictly, this is a semi-classical treatment where the classical Hamiltonian is 
replaced by the eigenenergies of the quantum system. In the following we show the validity of 
this semi-classical approach. We consider again the force operator $F_i= d{\mathbf v}/dt$, where ${\mathbf v}=\partial{\mathcal H}/\partial{\mathbf p}$ is the velocity.
The expectation value of the force operator in a  given quantum state $\psi_n$ is given as 
$\left\langle F_i\right\rangle=\langle\psi_n|\frac{d}{dt} (\partial_{p_i} {\mathcal H})|\psi_n\rangle$.
Explicitly,
\begin{eqnarray}
&&\langle\psi_n|\frac{d}{dt}(\partial_{p_i}{\mathcal H})|\psi_n\rangle=\frac{d}{dt}\partial_{p_i}\langle\psi_n|{\mathcal H}|\psi_n\rangle\nonumber\\
&&+\frac{d}{dt}\langle\partial_{p_i}\psi_n|{\mathcal H}|\psi_n\rangle+\frac{d}{dt}\langle\psi_n|{\mathcal H}|\partial_{p_i}\psi_n\nonumber\rangle\\
&&=\frac{d}{dt}\partial_{p_i}\epsilon_n+\frac{d}{dt}(\epsilon_n2Re\langle\psi_n|\partial_{p_i}\psi_n\rangle)\nonumber\\&&=\frac{d}{dt}(\partial_{p_i}\epsilon_n),
\end{eqnarray}
where in the last line we have used the fact that
$2Re\langle\psi_n|\partial_{p_i}\psi_n\rangle=\partial_{p_i}(\langle\psi_n|\psi_n\rangle)=0$. Therefore, the classical
derivation of force is actually equivalent to taking the expectation value of the force operator.

\end{document}